# A Comprehensive Survey of Upgradeable Smart Contract Patterns

Sajad Meisami
*Department of Computer Science*
*Illinois Institute of Technology*
Chicago,USA
smeisami@hawk.iit.edu

William Edward Bodell III
*Department of Computer Science*
*Illinois Institute of Technology*
Chicago,USA
wbodell@hawk.iit.edu

*Abstract*— **In this work, we provide a comprehensive survey of smart contract upgradability patterns using proxies. A primary characteristic of smart contracts on the Ethereum blockchain is that they are immutable – once implemented, no changes can be made. Taking human error into account, as well as technology improvements and newly discovered vulnerabilities, there has been a need to upgrade these smart contracts, which may hold enormous amounts of Ether and hence become the target of attacks. Several such attacks have caused tremendous losses in the past, as well as millions of dollars in Ether which has been locked away in broken contracts. Thus far we have collected many upgradable proxy patterns and studied their features to build a comprehensive catalog of patterns. We present a summary of these upgradable proxy patterns which we collected and studied. We scraped the source code for approximately 100k verified contracts from Etherscan.io, the most popular "block explorer" for Ethereum, out of which we extracted ~64k unique files - most containing multiple contracts. We have begun to automate the analysis of these contracts using the popular static analysis tool Slither, while at the same time implementing much more robust detection of upgradable proxies using this framework. Comparing the results of the original implementation to our own, we have found that approximately 70% of the contracts which were initially flagged as upgradeable proxies are false positives which we have eliminated.**

I. INTRODUCTION

*A. Smart Contracts*

Smart contracts, as first introduced with the genesis of the Ethereum blockchain, are typically short software programs that are stored on and executed by the platform's virtual computer, known as the Ethereum Virtual Machine (EVM) in the case of the original, best known and most used network. Considered immutable once they are deployed, smart contracts are unlike other forms of software in that they cannot easily be patched when bugs or vulnerabilities are discovered, and the network cannot be temporarily shut down for maintenance. Moreover, in recent years the cryptocurrency market has seen several dramatic surges in interest and price, with smart contracts gradually accumulating more and more of their users' wealth. With these concerns in mind, and also for the sake of enabling the iterative software development methodologies that have found success in other fields, a substantial amount of work has been put into developing security analysis tools for smart contracts, particularly for Ethereum.

For much the same reasons, a great deal work has been contributed by the Ethereum community to define upgradable smart contract patterns. Although there have been different approaches over the years, by the end of 2020, if not earlier, the community had largely settled on a preferred pattern, albeit one with many flavors. The proxy pattern for smart contract upgradability is really a family of patterns that all derive from the following recipe: a) the program is split into at least two parts, with upgradable logic contract(s) and an immutable storage contract, i.e. the proxy, b) the proxy contract stores the Ethereum address of the current logic implementation, which can be updated, and c) the proxy contract executes code from the logic contract whenever it receives a call to an unrecognized function, making use of the *delegatecall* low level EVM call which executes the delegate's code in the context of the caller contract.

*B. Slither*

Slither, one of the industry's more popular static analysis frameworks for smart contracts, assesses the security of source code written with Solidity. It uses a number of detectors written in Python to identify known vulnerabilities, as well as bad practices, with the collection of detectors growing as new vulnerabilities are reported. In addition to these detectors, Slither has added additional checkers for upgradable smart contracts, and more continue to be added as it is an open source project with an active community of contributors. However, in the course of testing these tools and inspecting the source code we discovered a number of shortcomings in the initial detection of upgradable contracts, specifically with the method in Slither's Contract class which determines if it is an upgradable proxy contract. We detail our findings with regard to false positives (FPs) and false negatives (FNs) resulting from the original algorithm in the following section, in the process providing some background as to why our modifications were necessary.

*C. Organization*

In section II we provide the necessary background information regarding smart contracts on Ethereum and the basics of upgradeable proxy patterns, as well as how Slither originally went about detecting them. In section III we present our modified algorithm, along with several examples we've encountered to demonstrate the different scenarios it has to handle. In section IV we compare the results of our improved algorithm to the original, and we also compare it with a tool provided by Etherscan, the most popular so-called "block explorer" for Ethereum, which detects proxy contracts and attempts to locate their implementations. Finally, we suggest further improvements to Slither's upgradable proxy contract detection which would better handle the growing variety of more sophisticated proxy patterns. As this research is very much a work in progress, this section serves as a road map for where we intend to go next.

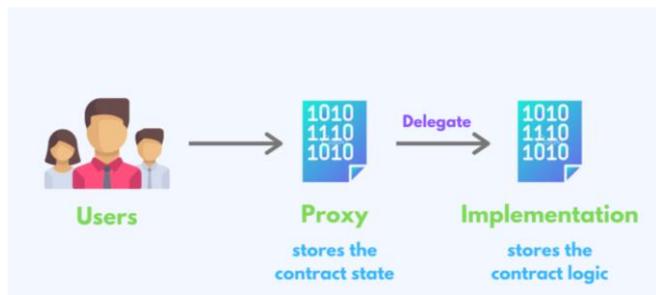

Fig. 1. Proxy Contract

## II. BACKGROUND

Smart contracts are simply programs stored on a blockchain that run when predetermined conditions are met. They typically are used to automate the execution of an agreement so that all participants can be immediately certain of the outcome, without any intermediary's involvement or time loss. A smart contract upgrade is an action that can arbitrarily change the code executed in an address while preserving storage and balance.

Most of these upgradability patterns for smart contracts depend on the *DELEGATECALL* opcode which is an EVM primitive. The way is to use a proxy contract with an interface where each method delegates to the implementation contract (which contains all the logic) with using *DELEGATECALL* opcode. Figure 1 shows how a basic proxy contract works. A delegate call is similar to a regular call, except that all code is executed in the context of the caller (proxy), not of the callee (implementation). Because of this, a transfer in the implementation contract's code will transfer the proxy's balance, and any reads or writes to the contract storage will read or write from the proxy's storage. This approach is better because the users only interact with the proxy contract and we can change the implementation contract while keeping the same proxy contract. In this way if we need to make any changes to the implementation contract methods, we would need to update the proxy contract's methods too (as the proxy contract has interface methods). Hence, users will need to change the proxy address. To solve this problem, we can use a fallback function in the proxy contract. The fallback function will execute on any request, redirecting the request to the implementation and returning the resulting value (using opcodes). Here the proxy contract does not have interface methods, only a fallback function, so there is no need to change the proxy address if contract methods are changed.

Slither was first published in August 2019, and in the subsequent years the open-source project has grown and adapted as Ethereum itself has been upgraded. With such a comprehensive tool and so many changes to the underlying protocol there are bound to be portions of the source code which are less commonly used and therefore more likely to break without it being noticed. This may be the case with Slither's upgradability related functions, which do appear to work as intended when the smart contract in question specifies an older version of *solc*, the Solidity compiler. We found that the tool failed to correctly identify upgradable proxy contracts using *solc versions >=0.6.0*, due to a change the Solidity developers made to the way inline assembly code is interpreted by the compiler. Beyond this breaking change, however, we also discovered a number of flaws in the algorithm that lead to both FPs and FNs.

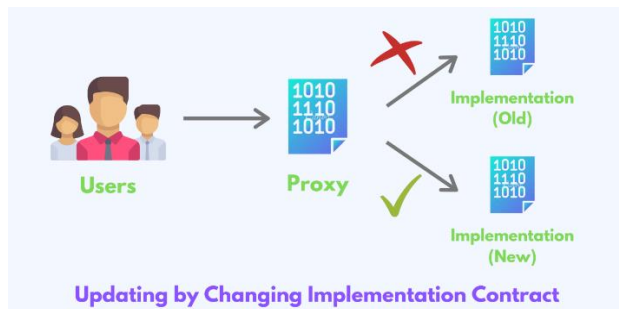

Fig. 2. Upgrading Implementation by proxy

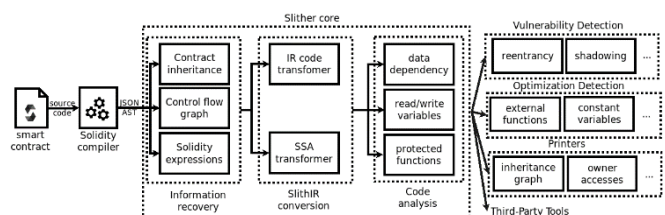

Fig. 3. Slither architecture

The most immediately noticeable flaw in the algorithm which would certainly lead to false positives is checking for the word "*Proxy*" in the name of the contract and returns true if found. Even if simply checking the contract's name were a valid method for determining whether it is a proxy, this would in no way confirm that the contract is an upgradable proxy. There are in fact many proxy contracts in the wild that cannot be upgraded by design; such minimal proxy contracts are typically used to reduce the cost of deploying many copies of a contract with somewhat long code, since the size of the contract factors into the gas cost of deploying it. On the other hand, there are numerous contracts out there which contain "*Proxy*" in their names yet are not themselves proxies. For instance, a common pattern is to use a "*ProxyAdmin*" contract to handle all upgrades, or a "*ProxyFactory*" which deploys new proxies. The unmodified algorithm would incorrectly flag both of these without ever looking for the most reliable indicator of a proxy, the low-level operation "*delegatecall*".

## III. UPGRADEABLE SMART CONTRACT PATTERNS

### A. Proxy Contracts Using Delegatecall

As mentioned before, most of these upgradable patterns depend on the *DELEGATECALL* opcode, so let's start with a brief overview of how it works. In a regular CALL from a contract A to a contract B, contract A sends a data payload to B. Contract B executes its code in response to this payload, potentially reading or writing from its own storage, and returns a response to A. While B executes its code, it can access information on the call itself, such as the *msg.sender*, which is set to A. However, on a *DELEGATECALL*, while the code executed is that of contract B, execution happens in the context of contract A. This means that any reads or writes to storage affect the storage of A, not B.

Delegate calls open the door to the proxy pattern and its many variants. Delegate calls open the door to the proxy pattern and its many variants. The proxy knows the implementation contract address, and delegates all calls it receives to it. Since the proxy uses a delegate call into the implementation, it is as if it were running the implementation's code as its own. It modifies its own storage and balance, and preserves the original *msg.sender* of the call.

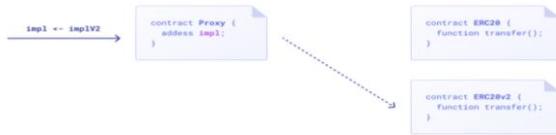

Fig. 4. Updating Implementation address in proxy

```
1.   contract AdminUpgradeableProxy {
2.     address implementation;
3.     address admin;

4.     fallback() external payable {
5.       implementation.delegatecall.value(msg.value
         )(msg.data);
6.     }

7.     function upgrade(address newImplementation)
         external {
         a.   require(msg.sender == admin);
         b.   implementation = newImplementation;
8.     }
9.   }
```
Fig. 5. Example code for updating Implementation address in proxy

Users always interact with the proxy, and are oblivious to the backing implementation contract.

Executing an upgrade is then straightforward. By changing the implementation address in the proxy, it is possible to change the code run upon every call to it, while the address the user interacts with is always the same. State is also preserved, since it is kept in the proxy's storage rather than the implementation contract.

The contract upgrade is usually done by a function that modifies the implementation contract. In some variants of the pattern, this function is coded into the Proxy directly, and restricted to be called only by an administrator. In Fig. 5 we have an exaple code which have a upgrade function to update Implimentation adress by a new one.

Due to the different models available, today there are several different types of proxy patterns that are used to upgrade smart contacts. We have tried to compile different types of upgradable proxy patterns in Table I and compare them so that we can find a suitable method for detecting smart contract upgradability. These proxy patterns are usually based on proposed EIPs (Ethereum Improvement Proposals).

*B. Avoiding Storage Clashes Between Proxy and Logic*

In all the proxy pattern variants, the proxy contract requires at least one state variable to hold the implementation contract address. By default, Solidity stores variables in the smart contract storage in order to the first variable declared goes to slot zero, the next to slot one, and so forth (mappings and dynamic-size arrays are exceptions to this rule). This means that, in the following proxy contract, the implementation will be saved to the storage slot zero. Now, what happens if we use that proxy combined with the seemingly innocuous implementation contract shown in Fig.6 Following Solidity storage layout rules, any calls to Implementation made through the proxy will store the first address variable in the storage slot zero. But keep in mind that, since we are using delegate calls, the storage affected will be that of the proxy, not the implementation contract. So calling into function in implementation would accidentally overwrite the proxy implementation address –something we definitely do not want to happen. To avoid this issue, the unstructured storage pattern was introduced. This pattern

```
1.   contract Box {
2.     address public value;
3.
4.     function setValue(address newValue)
         public {
5.       value = newValue;
6.     }
7.   }
```
Fig. 6. Example of an implementation contract with storage clash

mimics how Solidity handles mappings and dynamic-size arrays.

*C. Avoiding Function Selector Clashes*

All function calls in Ethereum are identified by the first 4 bytes of the data payload, which is known as the function selector. clashing between two functions occoures when two different functions with different names may end up having the same selector. it is perfectly possible for an implementation contract to have a function that has the same 4-byte identifier as the proxy's upgrade function. This could cause an admin to inadvertently upgrade a proxy to a random address while attempting to call a completely different function provided by the implementation. This issue can be solved either by appropriate tooling while developing upgradeable smart contracts, or at the proxies themselves. In particular, if the proxy is set up such that the admin can only call upgrade management functions, and all other users can only call functions of the implementation contract, clashes are not possible. Transparent proxy pattern aim is to solve this problem.

*D. Storage Layout Compatibility*

Contract upgrades introduce another challenge with regards to storage, in this case not between the proxy and the implementation, but between two different versions of the implementation. Let's suppose we have the following implementation contract deployed behind a proxy. A few months later, a new developer comes along and introduces some changes to this contract. As part of the new changes, they decide to sort the state variables alphabetically (just because they want to), and upgrade the contract in production.

Keep in mind the Solidity compiler decides to map variables to the contract storage based on the order in which the variables are declared. This means that, after the upgrade, the value of "number" is now in the slot assigned to "owner", and vice versa. This shows a major limitation of smart contract upgrades. while it's possible to arbitrarily change the code of a contract, only storage-compatible changes can be done to its state variables. Operations such as reordering variables, inserting new variables, changing the type of a variable, or even changing the inheritance chain of a contract can potentially break storage. The only safe change is to append state variables after any existing ones.

A pattern developed to address storage layout compatibility is the eternal storage pattern. This pattern uses the same strategy as unstructured storage, but for all variables of the implementation contract. This means that the implementation contract never declares any variables of its

## IV. CHALLENGES IN CONFIRMING PROXY UPGRADABILITY WITH SLITHER

As noted in the previous sections, Slither checks only name of proxy or detecting upgradable proxy pattern and it is not true way. Also, finding the *delegatecall* opcode in a contract's fallback function may be sufficient to determine that it is a proxy, but it is not enough to say that it is an *upgradable* proxy contract. For this reason, we can move most of the original algorithm's logic into the new *@property* definition shown in Fig.7. which we can call simply *is_proxy*. We check this property in the first line of the revised *is_upgradeable_proxy* algorithm in Fig.8. In addition to *self._is_proxy*, checking this also sets another property of type Variable called *self._delegates_to*, provided that it does find "*delegatecall*" and can identify its target. Currently this additional *self._delegates_to* property only returns the address variable which stores the implementation address, though in the future it may be desirable - and in some cases necessary - to return a Slither Contract object representing the implementation itself.

### A. Finding "delegatecall" in inline assembly for different Solidity versions

While testing the original algorithm we discovered that Slither parses contracts a bit differently depending on the Solidity version. While we have yet to fully understand the reasons for this, we did determine why the *node.inline_asm* property was missing for versions above 0.6.0. We fixed the issue by adding one line to the code in Listing 2, as mentioned in Section 2. Yet because of the change to how inline assembly is represented, it becomes more complicated to extract the target variable from the *delegatecall* operation. This is why we call the utility function *find_delegatecall_in_asm*, which is shown in Fig.9. This function searches the assembly code, which can be either as a Yul AST or simply a string, and also attempts to determine the target variable if it does find *delegatecall*.

However, we have found that it can be more difficult to accurately determine the target variable from the Yul AST. Thankfully, when using *Solc* versions >= 0.6.0 it is also possible to find a *delegatecall* and determine its target from a SlithIR CFG node object of type EXPRESSION. Hence, in Fig.7 serve as a backup which typically finds *self._delegates_to* if the prior two search methods do not yield results first. At any rate, *is_proxy* must return True and *delegates_to* must return a Variable in our revised *is_upgradeable_proxy* algorithm. If the target variable is constant, *is_upgradeable_proxy* will return False; otherwise, it proceeds to search the contract's functions, excluding any constructor and fallback, looking for the setter function which handles the upgrade logic.

### B. Finding the implementation setter in upgradable proxy contracts

Currently, our revised algorithm correctly identifies proxy contracts which contain logic for upgrading the implementation, for all Solidity versions that we have encountered thus far. It also avoids false positives that were previously caused either by finding the string "*Proxy*" in the contract's name, or by encountering the use of "*delegatecall*" in the fallback function of a proxy whose implementation address is immutable.

```
@property
def is_proxy(self) -> bool:
1.  if fallback_function is None:
2.      is_proxy = False
3.  for node in fallback_function.cfg_nodes:
4.      is_proxy, delegates_to =
            find_delegatecall_in_ir(node)
5.      if is_proxy and delegates_to is not None
6.          break
7.      if node.type == ASSEMBLY:
8.          is_proxy, delegates_to =
              find_delegatecall_in_asm(node.asm)
9.      elif node.type == EXPRESSION:
10.         is_proxy, delegates_to =
              find_delegatecall_in_exp_node(node)
11. return is_proxy
```
Fig. 7. New property is_proxy in Slither's Contract class, containing the logic from original Contract.is_upgradeable_proxy

```
@property
def is_upgradeable_proxy(self) -> bool:
1.  if not is_proxy or delegates_to is None:
2.      is_upgradeable_proxy = False
3.  if delegates_to.is_constant:
4.      is_upgradeable_proxy = False
5.  if delegates_to.contract != self:
6.      self.proxy_impl_setter =
          find_setter_in_contract(
            delegates_to.contract, delegates_to)
7.  else:
8.      self.proxy_impl_setter =
          find_setter_in_contract(self,
            delegates_to)
9.  if self.proxy_impl_setter is not None:
10.     self.is_upgradeable_proxy = True
11. if delegates_to.contract != self:
12.     self.proxy_impl_getter =
          find_getter_in_contract(
            delegates_to.contract, delegates_to)
13. else:
14.     self.proxy_impl_getter =
          find_getter_in_contract(self,
            delegates_to)
15. if self.proxy_impl_getter is not None:
16.     if self.proxy_impl_setter is not None:
17.         self.is_upgradeable_proxy = True
18.         return self.is_upgradeable_proxy
19.     else:
20.         return self.getter_return_nonconstant
21. else:
22.     return self.find_sload_from_ hardcoded_
          storage_slot(fallback_function)
```
Fig. 8. Overhauled property Contract.is_upgradeable_proxy which determines whether the implementation address can be updated

```
def find_delegatecall_in_asm(asm, parent_func:
1.  if "AST" in asm:
2.      // search Yul AST for "delegatecall"
3.      // if found, set self.is_proxy = True
4.      // and extract dest name as string
5.  else:
6.      asm_split = asm.split("\n")
7.      // search each line of assembly code for
8.      // "delegatecall", if found, set
9.      // self.is_proxy = True and extract dest
10.     // name as string
11. if self.is_proxy and dest is not None:
12.     delegates_to find_delegate_var_from_
          _name(dest, parent_func)
13. return is_proxy, delegates_to
```
Fig. 9. New helper function to search an assembly code block for "delegatecall" and extract the name of the destination variable

TABLE I.   UPGRADABLE PROXY PATTERNS

| Proxy Pattern Name | Brief Description | Documentation |
|---|---|---|
| **Inherited Storage** | Both the proxy and the logic contract inherit the same storage structure to ensure that both adhere to storing the necessary proxy state variables. | https://blog.openzeppelin.com/proxy-patterns/ https://github.com/OpenZeppelin/openzeppelin-labs/tree/master/upgradeability_using_inherited_storage/ |
| **Eternal Storage** | Developed to address storage layout compatibility across upgrades. The proxy and implementation contracts never declare any variables of their own, but rather store them in a mapping, which causes Solidity to save them in arbitrary positions of storage, based on their assigned names. | https://blog.openzeppelin.com/smart-contract-upgradeability-using-eternal-storage/ |
| **Unstructured Storage EIP-1967 OpenZeppelin** | This pattern mimics how Solidity handles mappings and dynamic-size arrays to avoid proxy storage clashes: it stores the implementation address variable not in the first slots, but in an arbitrary slot in storage. Given the addressable storage of a contract is 2^256 in size, chances of a clash are effectively zero. Creation: 2019-04-24 | https://blog.openzeppelin.com/upgradeability-using-unstructured-storage/ https://eips.ethereum.org/EIPS/eip-1967 |
| **EIP-1822 Universal Upgradeable Proxy Standard** | These standard places upgrade logic in the implementation contract instead of the proxy itself. UUPS proposes that all implementation contracts extend from a base proxiable contract. Storage of the implementation address is handled as in unstructured storage, i.e at a storage slot equal to keccak256("PROXIABLE") Creation: 2019-03-04 | https://eips.ethereum.org/EIPS/eip-1822 |
| **(Admin only) Transparent Proxy** | Another extension of unstructured storage, in which the admin can only call upgrade management functions, and all other users can only call functions of the implementation contract, so selector clashes are not possible. (Disadvantage: High Gas cost due to the extra load from memory to check admin) | https://blog.openzeppelin.com/the-transparent-proxy-pattern/ |
| **EIP-1538 Transparent Proxy** | In this version, instead of storing a single implementation address, the proxy stores a mapping from function selector to implementation address. This EIP was withdrawn and replaced by EIP-2535. Creation: 2018-10-31 | https://eips.ethereum.org/EIPS/eip-1538 |
| **EIP-2535 Diamonds Proxy** | Replaces EIP-1538 In this version, instead of storing a single implementation address, the proxy stores a mapping from function selector to implementation address. Uses terminology from the diamond industry: - logic contracts are called Facets - a Loupe is a Facet that gives visibility to the list of Facets for the Diamond - function 'Facets' can be added, removed, or modified using diamondCut() Creation: 2020-02-22 | https://eips.ethereum.org/EIPS/eip-2535 |
| **Beacons Proxy** | Beacons enable multiple proxies per implementation, with the ability to upgrade them all at once. Each proxy holds the address not to its implementation contract, but to a beacon which, in turn, holds the address of the implementation. | https://blog.dharma.io/why-smart-wallets-should-catch-your-interest/ |
| **Registry Proxy** | Like a beacon proxy, it must query a Registry contract to get its implementation address. However, the registry maps a version number to an associated implementation address. Also, in many cases the registry can serve only one proxy. Not to be confused with ProxyRegistry as used by Maker, which maps a user address to a DSProxy. | https://medium.com/@blockchain101/the-basics-of-upgradable-proxy-contracts-in-ethereum-479b5d3363d6 |
| **EIP-1167 Minimal Proxy** | It reduces deployment costs, when multiple instances of a contract are needed. Deploying several copies of a large contract can be very expensive in terms of gas costs, so it's more cost-effective to deploy a single copy to act as an implementation contract, and spawn multiple proxies backed by it. These proxies do not need to be upgraded, they do not need any storage or management functions, making them dead-simple. Creation: 2018-06-22 | https://eips.ethereum.org/EIPS/eip-1167 |
| **EIP-897 DelegateProxy** | This standard proposes a set of interfaces for proxies to signal how they work and what their main implementation is. Creation: 2018-02-21 | https://eips.ethereum.org/EIPS/eip-897 |

## V. CONCLUSION AND NEXT STEPS

In this work, we provide a comprehensive survey of smart contract upgradability patterns using proxies. A primary characteristic of smart contracts on the Ethereum blockchain is that they are immutable – once implemented, no changes can be made. Thus far we have collected many upgradable proxy patterns and studied their features to build a comprehensive catalog of patterns. We present a summary of these upgradable proxy patterns which we collected and studied. We have begun to automate the analysis of these contracts using the popular static analysis tool Slither, while at the same time implementing much more robust detection of upgradable proxies using this framework. Comparing the results of the original implementation to our own, we have found that approximately 70% of the contracts which were initially flagged as upgradeable proxies are false positives which we have eliminated.

Thankfully, Slither's Contract class already provides access to a contract's "compilation unit", i.e. all of the contracts and libraries that are included or imported in the source code. Thus we should be able to identify the "manager" contract in question. This should make confirming upgradability trivial, provided that the source code for this additional contract is available as it is in this case. However, there is no guarantee that this will always be the case. Another example of an upgradable proxy contract that results in a false negative with our current algorithm follows the pattern proposed by EIP-1822: Universal Upgradeable Proxy Standard. Similar to EIP-1967, this standard proposes a standard storage slot for storing the implementation address with the proxy. Yet unlike the more common pattern, with EIP-1822 the function used for updating the address resides in the logic contract itself, and there is not necessarily any indication in the proxy contract's source code as to where it may be located.

In this case and possibly others it may be necessary to adopt another approach, perhaps looking outside of Slither's static analysis environment for answers. One alternative is make use of or adapt a tool that was added fairly to Etherscan, the most popular so-called "block explorer" for Ethereum. The tool flags possible proxy contracts in which it can find a delegatecall opcode sequence, and attempts to locate their implementations using several heuristics, such as by checking for addresses stored at known common storage slots. However, we have observed that Etherscan's proxy frequently fails to find the implementation address. Furthermore in many cases in which it does find the address, the current version of the logic contract may not have uploaded to Etherscan for verification. As such, there is room for improvement on this front as well.

## REFERENCES

[1] https://docs.soliditylang.org/en/v0.8.9/yul.html
[2] https://blog.openzeppelin.com/proxy-patterns/
[3] Bodell III, William E., Sajad Meisami, and Yue Duan. "Proxy Hunting: Understanding and Characterizing Proxy-based Upgradeable Smart Contracts in Blockchains."
[4] Meisami, Sajad, Mohammad Beheshti-Atashgah, and Mohammad Reza Aref. "Using blockchain to achieve decentralized privacy in IoT healthcare." *arXiv preprint arXiv:2109.14812* (2021).
[5] Meisami, Sajad, Sadaf Meisami, Melina Yousefi, and Mohammad Reza Aref. "Combining Blockchain and IOT for Decentralized Healthcare Data Management." *arXiv preprint arXiv:2304.00127* (2023).
[6] Duan, Yue, Xin Zhao, Yu Pan, Shucheng Li, Minghao Li, Fengyuan Xu, and Mu Zhang. "Towards Automated Safety Vetting of Smart Contracts in Decentralized Applications." In *Proceedings of the 2022 ACM SIGSAC Conference on Computer and Communications Security*, pp. 921-935. 2022.
[7] A. Kosba, A. Miller, E. Shi, Z. Wen and C. Papamanthou, "Hawk: The Blockchain Model of Cryptography and Privacy-Preserving Smar Contracts," in *2016 IEEE Symposium on Security and Privacy (SP)*, San Jose, CA, USA, August 2016.
[8] *Blockchain for Financial Services*. Accessed: Mar. 25, 2018. [Online]. Available: https://www.ibm.com/blockchain/financial-services
[9] https://github.com/OpenZeppelin/openzeppelin-labs/tree/master/upgradeability_using_inherited_storage/.
[10] https://blog.openzeppelin.com/smart-contract-upgradeability-using-eternal-storage/
[11] https://blog.openzeppelin.com/upgradeability-using-unstructured-storage/
[12] https://eips.ethereum.org/EIPS/eip-1967
[13] https://eips.ethereum.org/EIPS/eip-1822
[14] https://blog.openzeppelin.com/the-transparent-proxy-pattern/
[15] https://eips.ethereum.org/EIPS/eip-1538
[16] https://eips.ethereum.org/EIPS/eip-2535
[17] https://blog.dharma.io/why-smart-wallets-should-catch-your-interest/
[18] Liu, Q.; Kosarirad, H.; Meisami, S.; Alnowibet, K.A.; Hoshyar, A.N. An Optimal Scheduling Method in IoT-Fog-Cloud Network Using Combination of Aquila Optimizer and African Vultures Optimization. *Processes* **2023**, *11*, x.
[19] https://medium.com/@blockchain101/the-basics-of-upgradable-proxy-contracts-in-ethereum-479b5d3363d6
[20] https://eips.ethereum.org/EIPS/eip-1167
[21] https://eips.ethereum.org/EIPS/eip-897
[22] https://gist.github.com/Arachnid/4ca9da48d51e23e5cfe0f0e14dd6318f
[23] https://blog.openzeppelin.com/proxy-patterns/
[24] https://blog.openzeppelin.com/smart-contract-upgradeability-using-eternal-storage/
[25] https://blog.openzeppelin.com/towards-frictionless-upgradeability/
[26] https://blog.openzeppelin.com/the-transparent-proxy-pattern/
[27] https://docs.openzeppelin.com/upgrades-plugins/
[28] https://blog.indorse.io/ethereum-upgradeable-smart-contract-strategies-456350d0557c
[29] https://medium.com/coinmonks/summary-of-ethereum-upgradeable-smart-contract-r-d-part-2-2020-db141af915a0
[30] https://blog.gnosis.pm/solidity-delegateproxy-contracts-e09957d0f201
[31] https://medium.com/@0age/the-promise-and-the-peril-of-metamorphic-contracts-9eb8b8413c5e
[32] https://blog.dharma.io/why-smart-wallets-should-catch-your-interest/
[33] Meisami, Sajad, Mohammad Beheshti-Atashgah, and Mohammad Reza Aref. "Using Blockchain to Achieve Decentralized Privacy In IoT Healthcare." Cryptology ePrint Archive (2021).
[34] https://dev.to/mudgen/understanding-diamonds-on-ethereum-1fb

[35] https://blog.openzeppelin.com/deep-dive-into-the-minimal-proxy-contract/

[36] https://blog.openzeppelin.com/the-state-of-smart-contract-upgrades/#transparent-proxies